\documentclass[12pt]{iopart}

\input epsf

\begin{document}
\jl{2}

\title{Bose-Einstein condensation in a rotating anisotropic TOP trap}
\author{J.\thinspace Arlt, O.\thinspace Marag\`o, E.\thinspace Hodby,
S.A.\thinspace Hopkins, G.\thinspace Hechenblaikner,
S.\thinspace Webster and C.J.\thinspace Foot.}
\address{Clarendon Laboratory, University of Oxford, Parks Road, Oxford, OX1 3PU, UK}
\date{\today}

\begin{abstract}
We describe the construction and operation of a time-orbiting
potential trap (TOP trap) that has different oscillation
frequencies along its three principal axes. These axes can be
rotated and we have observed Bose-Einstein condensates of
$^{87}$Rb with a rotating ellipsoidal shape. Under these
conditions it has been predicted that quantized vortices form and
are stable.
\end{abstract}


\section{Introduction}
The first experimental observations of Bose-Einstein condensation
in alkali metals in 1995 \cite{Anderson, Davis} and subsequent
experiments have led to much progress, both theorerical and
experimental, in understanding the physics of these condensed
systems (for a review see \cite{Varenna}). Initial experiments
focussed on the macroscopic properties of this new state of
matter, such as collective excitations \cite{Jin, Mewes}. Later
experiments on atom lasers \cite{Laser, Sodium} and the production
of multi-component condensates \cite{Component, Compo} have
focussed on the coherence properties of BEC's. In particular the
description of a condensate by a single wavefunction with a
definite phase has been beautifully demonstrated by matter-wave
interference experiments \cite{Interference, Interfere}.
Nevertheless, in contrast to earlier work on liquid helium
\cite{Tilley}, the spectacular phenomena of superfluidity have yet
to be comprehensively demonstrated and studied in the alkali metal
condensates e.g. non-viscous flow in narrow channels, quantised
vortices and persistent currents. Some of these phenomena are a
consequence of the irrotational nature of superfluid flow, which
in turn stems from the existence of the single-valued
wavefunction.  Recently much theoretical effort, by various groups
\cite{Theory, Stringari, Fetter, Lundh, Bolda}, has been devoted
to possible methods of formation and observation of quantised
vortices and persistent currents, which are associated with
rotation of the condensate. It has been shown that rotating an
anisotropic trap should create vortices in a similar manner to
those observed in liquid helium \cite{Tilley}.

In contrast to the time-averaged orbiting potential (TOP) trap
used in this work, it is difficult to create a rotating potential
using a magnetostatic trap such as the Ioffe-Pritchard
\cite{Ioffe} configuration. However it may be possible to study
superfluidity using other methods that have recently been
suggested: stirring a trap with light forces \cite{Stir},
imprinting a $2\pi$ phase-winding with light beams
\cite{Imprinting} and manufacturing a vortical wavefunction in a
spinor superposition state \cite{Vortex}. Only the latter method
has been successful so far; the challenge of observing a vortex
for a condensate in a single internal state has yet to be met.

However these alternative methods cannot directly test some of the
interesting theoretical predictions which have recently been made
for states in a rotating trap which have non-integer values of
angular momentum (in units of Plancks constant) per particle
\cite{Angular}. The method presented here allows rotation without
changing the ellipticity of the potential even for large
ellipticities. These features cannot be achieved in the modulation
method \cite{Jin} used to excite $m=2$ condensate excitations.

\section{Experimental arrangement and BEC in TOP trap}

We use a combination of laser cooling, magnetic trapping and
evaporative cooling to obtain a $^{87}$Rb condensate in the
$F=2,m_F=2$ state in a similar manner to the first observation of
BEC \cite{Anderson}. The experimental arrangement is similar to
those described elsewhere \cite{Petrich, Wilson}, and only
important features of our setup are given below.\\ The light for
our experiment comes from two external cavity diode lasers, {\it
master} and {\it repumper}, operating close to the frequency of
the rubidium D2-transitions. The light from the master laser is
injected and amplified by a tapered semiconductor amplifier chip
in a master oscillator power amplifier (MOPA) \cite{Wilson2}
configuration which provides $\sim 500$ mW of output. The light is
fed to two magneto-optical traps (MOT) in a double trap
configuration \cite{double} to preserve a good vacuum in the
experimental region while obtaining high loading rates.

A novel feature is the use of a pyramidal configuration of mirrors
to form the collection MOT \cite{Arlt2}. A large number of
rubidium atoms are emitted from a getter source \cite{Getter} in
this region and then continuously cooled and transferred down a
$30 cm$ pipe of $15 mm$ inner diameter, into the region of higher
vacuum where they are captured in the second MOT. Magnetic
guiding, in the transfer pipe, increases the flux of atoms to the
second MOT by a factor of $2$ \cite{double}. The second MOT is a
standard 6-beam MOT configuration used to capture and store the
atoms. In order to obtain large numbers of atoms we found that it
was beneficial to keep the small ($\approx 2G$) time-varying
magnetic field used in the TOP trap switched on during loading
into the second MOT. The Zeeman shift of this field causes atoms
near the centre to be detuned from resonance, giving a partial
dark MOT \cite{Kasevich}.\\ After collecting $5 \times 10^8$ atoms
in the second MOT they are compressed, cooled to a temperature of
$30\mu K$ in an optical molasses and then optically pumped into
the $F=2, m_F=2$ state. The quadrupole field is created by two
coils in anti-Helmholtz configuration with $500$ turns each,
operating at a maximum current of $10A$. The $7 kHz$ rotating
field is created by two sets of Helmholtz coils with 5 turns in
each coil. The current to each set of Helmholtz coils is supplied
via matching transformers from a $600W$ audio amplifier.

\begin{figure}
\begin{center}\mbox{ \epsfxsize 5in\epsfbox{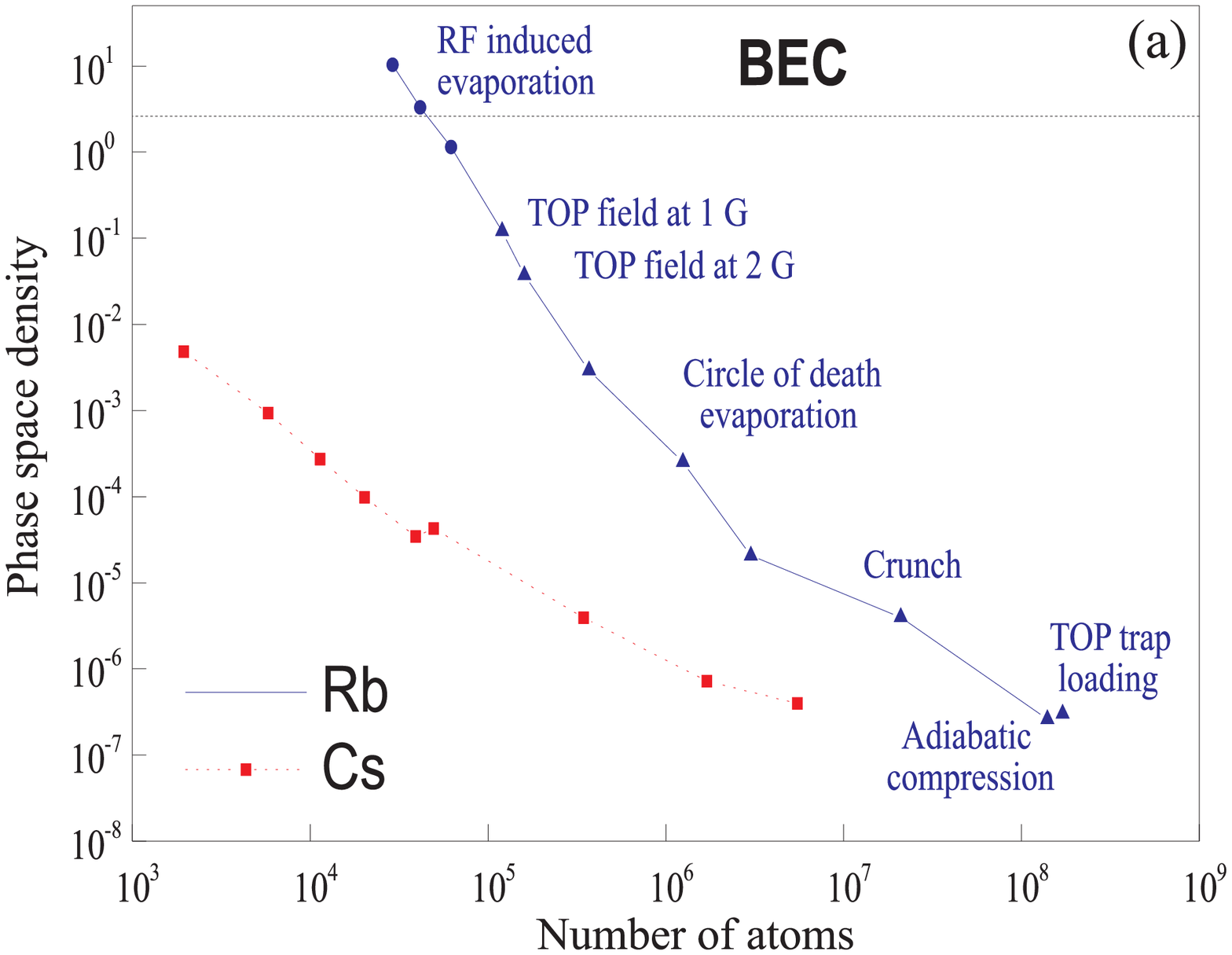}}\end{center}
\begin{center}\mbox{ \epsfxsize 5in\epsfbox{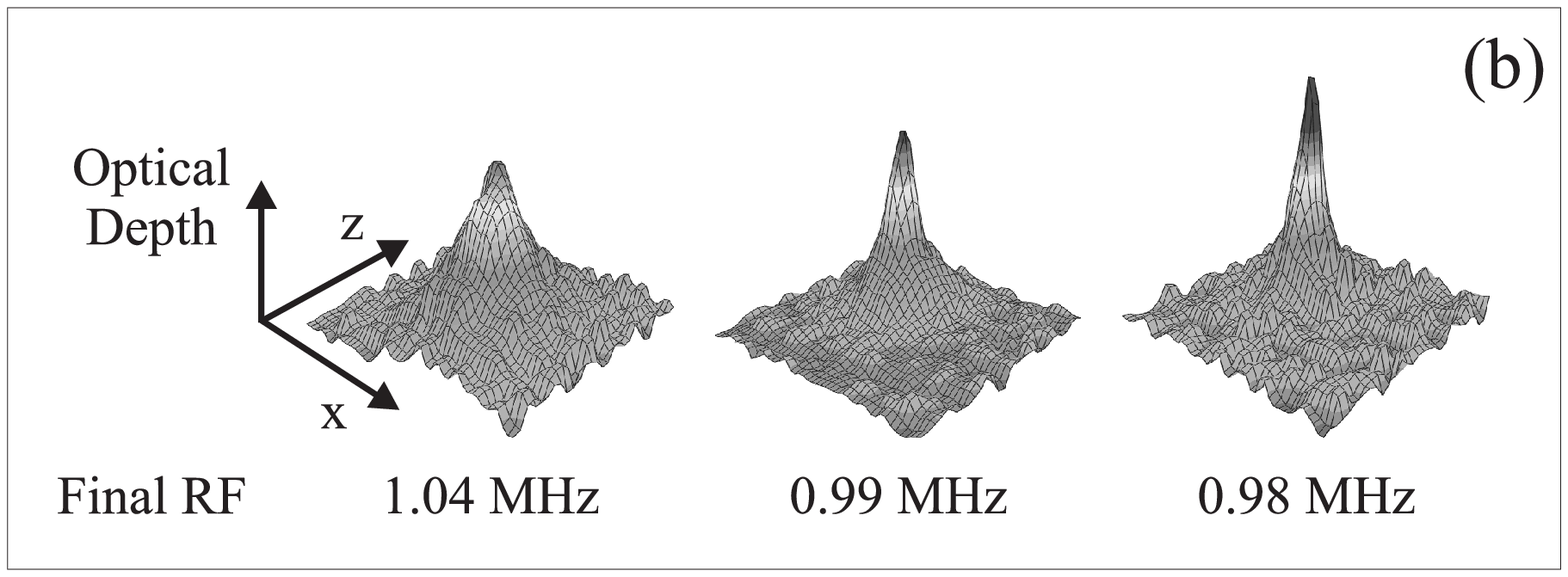}}\end{center}
\caption{(a) Plot of phase space density vs. number of atoms
during the evaporative cooling ramp. We compare the evaporative
cooling efficiency for $^{87}$Rb and and $^{133}$Cs in two
identical magnetic traps. "Crunch" refers to the combined
evaporation and compression ramp. (b) BEC formation in a trap with
$1G$ bias field. The images were taken after $7ms$ TOF using
absorption imaging.} \label{phaseplot}
\end{figure}

Once the atoms are loaded into a mode-matched TOP trap ($25G$ bias
field and $65G/cm$ radial gradient) we use the following procedure
to obtain BEC (see also Fig.\ref{phaseplot} (a)). Firstly the TOP
and quadrupole fields are adiabatically ramped to $40G$ and
$100G/cm$ in $2s$ to increase the collision rate. In a combined
evaporation and compression ramp the radial quadrupole field
gradient continues to rise to $194G/cm$ while the TOP field
decreases to $20G$ in $20s$. Now the TOP field itself is used to
evaporate the hottest atoms by cutting to $1G$ in $28s$ obtaining
a final radial trap frequency of $175 Hz$\footnote{The trapping
frequencies were measured by exciting 'dipole' oscillations of a
small thermal cloud with a sudden change of the bias magnetic
field.}. Finally a $5s$ radio frequency cut, finishing $60\%$ of
the way in from the radius of the locus of $B=0$
($\nu_{final}=0.98 MHz$), is used to obtain a pure condensate of
$\sim 3\times10^4$ atoms. In Fig.\ref{phaseplot}(a) it is possible
to follow our path to BEC for $^{87}$Rb in a phase space diagram
and compare it with the case of $^{133}$Cs \cite{Webster} that has
also been studied in a parallel experiment (using the same
magnetic trap) at our lab \cite{Arlt}.\\ The pictures in
Fig.\ref{phaseplot}(b) show the emergence of the pure condensate
as the depth of the final RF cut is increased. We observe the
sudden increase of optical depth and the reversal of shape
characteristic for a BEC in time of flight (TOF) imaging. A laser
pulse of $10\mu s$, synchronised with the TOP field and resonant
with the $F=2, m_F=2\rightarrow F^{\prime}=3, m_F^{\prime}=3$
transition, is used for absoption imaging of the cloud of atoms in
the xz-plane.

\section{Elliptical TOP potential}

The invention of the TOP trap \cite{Petrich} lead to the first
observation of BEC \cite{Anderson} in a dilute alkali vapour of
rubidium, and this type of trap has subsequently also been used by
other groups to trap caesium \cite{Arlt, Dalibard} and sodium
\cite{Sodium}. The TOP trap is formed by combining a spherical
quadrupole field, which can be written as

\begin{center}
\begin{equation}
\mathbf{B}_Q=B_Q^{\prime} \left( x{\bf e}_x +y{\bf e}_y -2z{\bf e}_z \right)
\end{equation}
\end{center}

\noindent near the origin ($B_Q^{\prime }$ is the gradient), with a rotating field in the xy-plane

\begin{center}
\begin{equation}
\label{Bellipse} \mathbf{B}= B_{x} \cos \omega_0t\ {\bf e}_x + B_{y} \sin \omega_0t\ {\bf e}_y
\end{equation}
\end{center}

\noindent The rotation frequency $\omega _0$ is chosen to be
greater than the oscillation frequencies of atoms in the trap
$\omega_x,\omega_y$ and $\omega_z$, but less than the Larmor
frequency so that atoms do not change $m_F$-state. The TOP trap is
usually operated in the axisymmetric configuration where the
amplitudes of the fields in the two directions are equal,
$B_{x}=B_{y}$, and the trapping frequencies are in the ratios
$\omega_x=\omega_y=\omega _z/\sqrt{8}$ \footnote{A non
axisymmetric TOP trap has been used to obtain BEC in Na
\cite{Sodium}.}. However it is numerically straightforward to
calculate the potential in the more general case where the ratio
of the magnitudes of the rotating fields is $B_{x}/B_{y}=E$. The
TOP potential is the time-average of

\begin{center}
\begin{equation}
U(x,y,z,t)=\mu B_Q^{\prime} | ( x+E r_0\cos \omega_0t ) {\bf e}_x+(y+r_0\sin \omega _0t) {\bf e}_y-2z {\bf e}_z | \label{TOP}
\end{equation}
\end{center}

\noindent where $\mu$ is the magnetic moment of the atoms and
$r_0=B_{y}/B_Q^{\prime }$ is the distance from the centre of
rotation at which the locus $B=0$ intersects the y-axis. For $E=1$
this locus is a circle of radius $r_0$ in the $xy$-plane, and in
general it is an ellipse of eccentricity $E$ in the $xy$ plane described
by the position vector

\begin{center}
\begin{equation}
\label{ellipseeq} \mathbf{r}=r_0 \left( E \cos \omega _0t\ {\bf e}_x+ \sin \omega _0t\ {\bf e}_y \right)
\end{equation}
\end{center}

\begin{figure}
\begin{center}\mbox{ \epsfxsize 5in\epsfbox{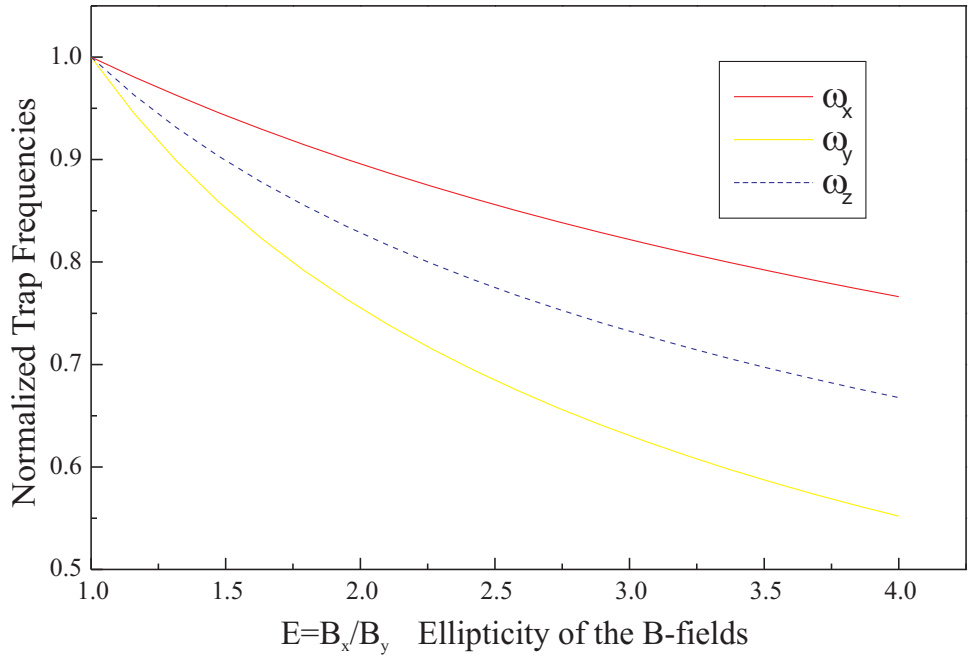}}\end{center}
\caption{Plot of the trapping frequencies in an elliptical TOP
trap as a function of the ellipticity $E=B_x/B_y$ where $B_x$ is
varying with $E$ for fixed values of $B_y$ and the quadrupole
gradient. The frequencies are plotted as fractions of their values
for $B_x=B_y$. The dotted line ($\omega _z$) remains approximately
equal to the mean of the other frequencies
$(\omega_x+\omega_y)/2$.} \label{Frequency}
\end{figure}

Without loss of generality we can consider $E \geq 1$ so that the
order of the numerically calculated trapping frequencies, shown in
Fig.\ref{Frequency}, is $\omega _x\leq\omega _y<\omega _z$. The
harmonic approximation to the TOP potential

\begin{center}
\begin{equation}
 \overline{U}=\frac12 m\left( \omega _x^2\,x^2+\omega_y^2\,y^2+\omega _z^2\,z^2\right)
\end{equation}
\end{center}

\noindent is accurate at distances small compared to $r_{0}$. We
are mainly interested in the deformation of the time-averaged
potential in the $xy$-plane, characterised by the parameter
$\omega _y/\omega _x=e\geq 1$. Contours of constant energy in the
$xy$-plane are ellipses given by

\begin{center}
\begin{equation}
x^2+e^2y^2=$constant$
\end{equation}
\end{center}

\noindent whose major axes are in the same direction as the major
axis of the ellipse in Eq. \ref{TOP}. Note however that the
eccentricity of the TOP is much less than $E$ and that the field
in the $x$-direction reduces all the trap frequencies, not just
$\omega_x$. Thus it is not possible to reach very large
deformations of the TOP potential. For small deformations the
ellipticity in the trap frequency varies as $de/dE=1/4$.

\section{Rotation of the TOP}

A rotation of an angle $\phi$ is mathematically obtained by a
transformation of coordinates expressed by the matrix

\begin{center}
\begin{equation}
\left(
 \begin{array}{cc}
  \cos\phi & \sin\phi \\ -\sin\phi & \cos\phi
 \end{array}
\right).
\end{equation}
\end{center}

\noindent Applying this with a time varying angle $\phi =\Omega t$
to Eq.\ref{ellipseeq} gives the position vector of a rotating
ellipse

\begin{center}
\begin{equation}
\begin{array}{lll}
 {\bf r}^{\prime} &=& r_0 \left( E \cos \Omega t\cos \omega _0t+\sin \Omega t\sin \omega_0t \right) {\bf e}_x + \\
 & & r_0 \left( -E \sin \Omega t\cos \omega_0t+\cos \Omega t\sin \omega _0t \right) {\bf e}_y \label{xyfields}.
\end{array}
\end{equation}
\end{center}

\begin{figure}
\begin{center}\mbox{ \epsfxsize 5in\epsfbox{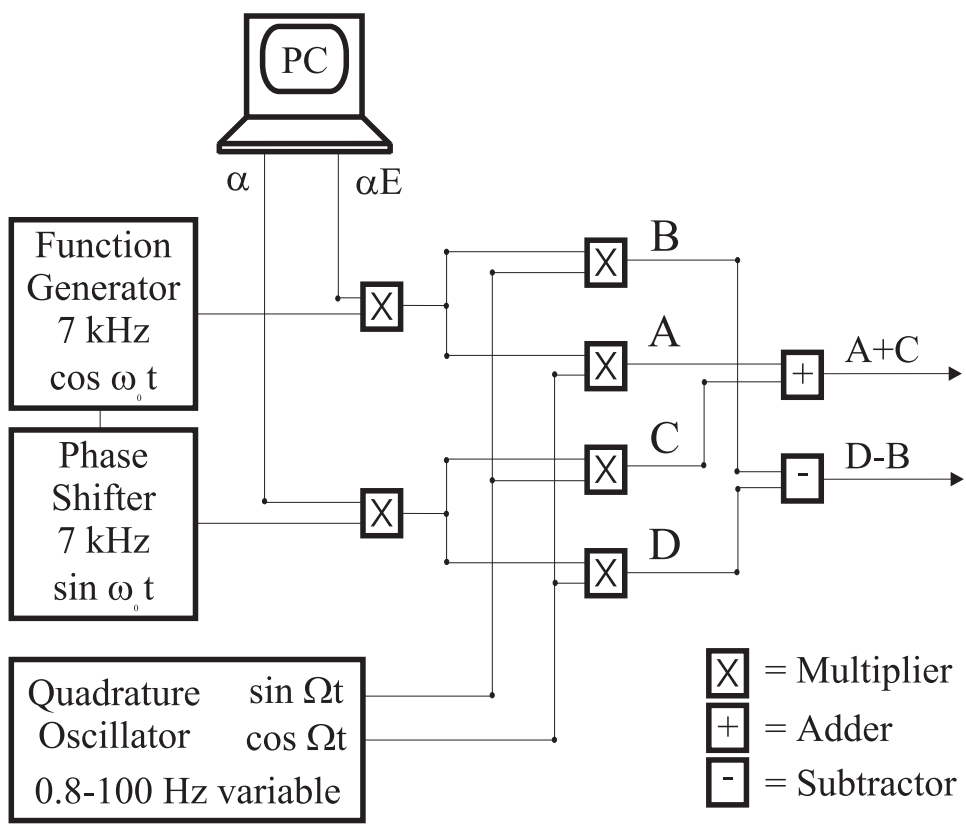}}\end{center}
\caption{Schematic of the circuit for modulating the sine and
cosine waves driving the TOP coils.} \label{Circuit}
\end{figure}

\noindent When $E=1$, this expression reduces to $\sin(\omega
_0-\Omega)t$ and $\cos(\omega _0-\Omega)t$, which corresponds to
the usual TOP configuration with a very slightly different
frequency, since $\Omega \ll \omega _0$. (e.g. $\Omega \leq 2\pi
\times 100Hz$ whereas $\omega _0=2\pi \times 7 kHz$ in these
experiments). To make the locus of $B=0$ a rotating ellipse as
described by Eq.\ref{xyfields}, we need to apply magnetic fields
in the x and y directions which vary in time in the way described
in these equations. This is achieved using the circuit shown
schematically in Fig.\ref{Circuit}, starting with four signals
proportional to $E \cos \omega _0t$, $\sin \omega _0t$, $\cos
\Omega t$ and $\sin \Omega t$. These are then multiplied in pair
by four voltage multiplier to give the four product voltages.

\begin{center}
\begin{equation}
\begin{array}{lll}
A &=&\cos \Omega t\; E\cos \omega _0t \\ C &=&\sin \Omega t\; \sin
\omega _0t \\ B &=&\sin \Omega t\; E\cos \omega _0t \\ D &=&\cos
\Omega t\; \sin \omega _0t\
\end{array}
\end{equation}
\end{center}

\noindent These voltages are then added (or subtracted) by simple
operational amplifier circuits to yield final signals

\begin{center}
\begin{equation}
\begin{array}{lll}
B_x &\propto &A+C \\ B_y &\propto &D-B,
\end{array}
\end{equation}
\end{center}

\noindent which are amplified by the audio amplifier and fed to
the coils producing the time-varying fields along x and y.\\ It is
important to ensure that any high frequency noise introduced in
the multiplication and summing stages is filtered out before the
amplifier otherwise it leads to a large decrease in the lifetime
of the trapped atoms. This 'noise-induced' loss becomes more
severe at lower values of the rotating bias field, and our present
system cannot be operated below $1 G$.

\section{BEC in the rotating anisotropic TOP trap}
A simple description of the behaviour of trapped atoms in the
elliptical rotating trap, is obtained from the classical equations
of motion for anisotropic harmonic oscillation of a particle in a
frame rotating at angular frequency $\Omega $ about the z-axis

\begin{equation}
\begin{array}{lll}
\ddot{x} &=&-\left( \omega _x^2-\Omega ^2\right) x-2\Omega \
\dot{y} \\ \ddot{y} &=&-\left( \omega _y^2-\Omega ^2\right)
y+2\Omega \ \dot{x} \\ \ddot{z} &=&-\omega _z^2\ z
\end{array}
\end{equation}

\noindent These equations are discussed in many mechanics
textbooks e.g. in the book by Lamb \cite{Lamb}, he shows that the
two equations for the xy motion describe Blackburn's pendulum, a
pendulum with different effective lengths in two orthogonal
directions on a turntable. The oscillation frequencies in the $x$
and $y$ directions are reduced by the effect of the centripetal
acceleration outwards and the $x$ and $y$ equations are coupled by
the Coriolis force $\mathbf{\Omega \times \dot{r}}$. One obvious
feature of these equations is that the motion becomes unstable
when the rotation is equal to the lowest trap oscillation
frequency, $\Omega=\omega _x$.

\begin{figure}
\begin{center}\mbox{ \epsfxsize 5in\epsfbox{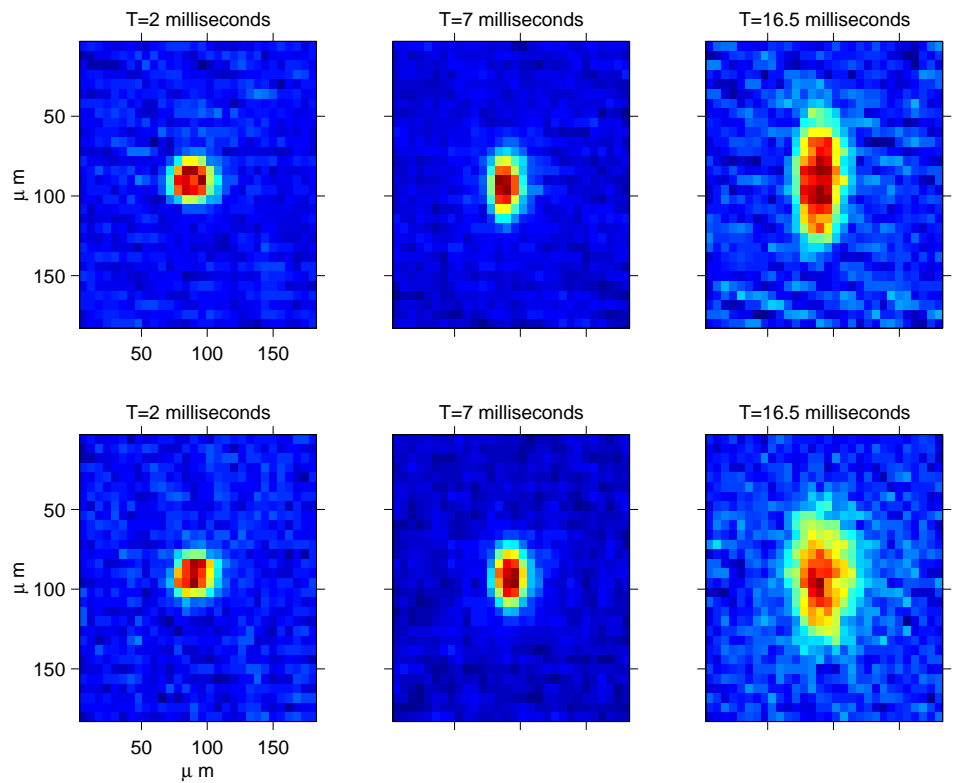}}\end{center}
\caption{Pictures of BECs at different times of flight after being
released from the elliptical rotating TOP trap. The top row shows
BECs released with their x axes parallel to the imaging axes. The
bottom row shows BECs released with their y axes parallel to the
imaging axes.} \label{BECpic}
\end{figure}

\begin{figure}
\begin{center}\mbox{ \epsfxsize 5in\epsfbox{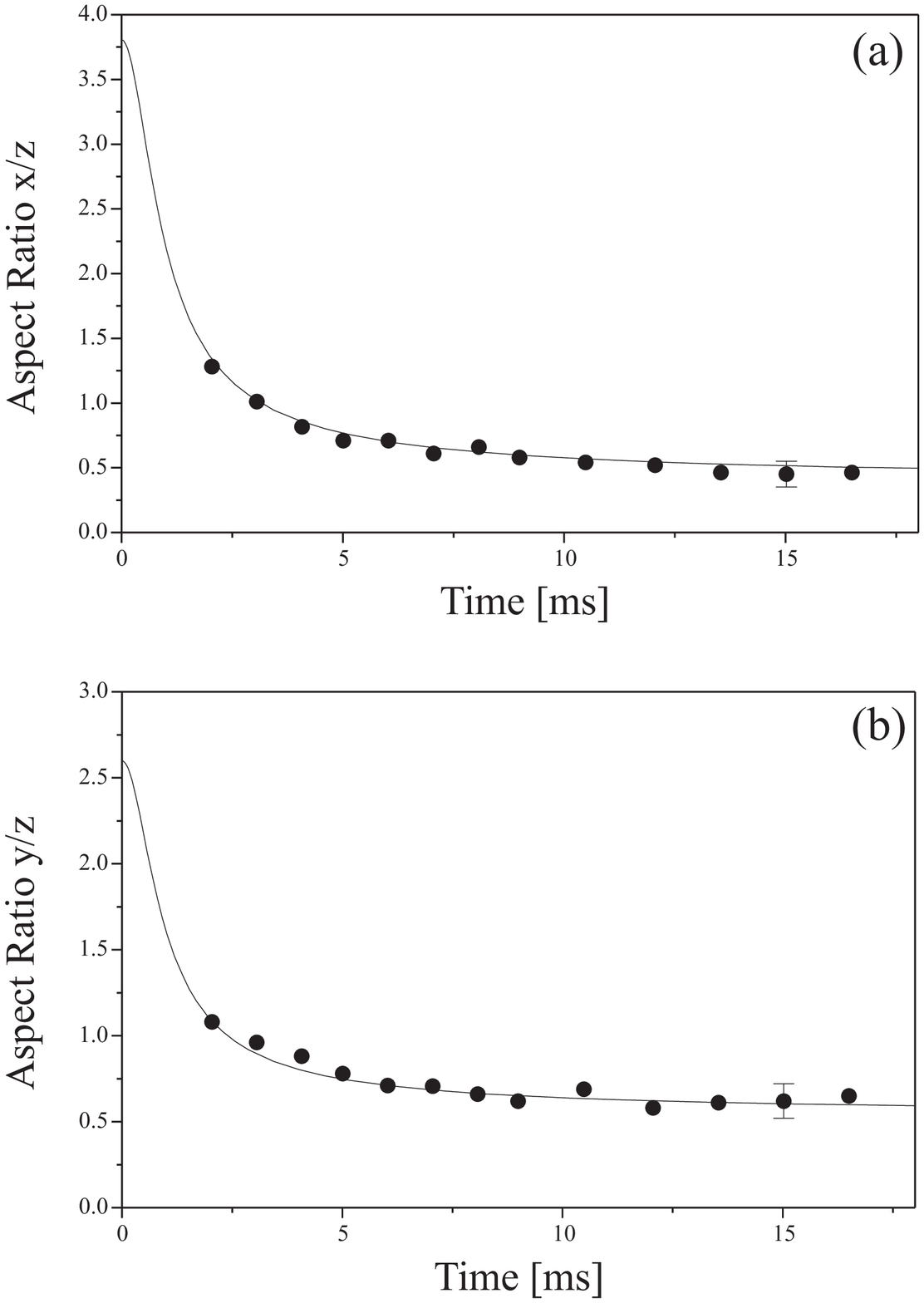}}\end{center}
\caption{The expanison of the BEC is shown as a function of time.
The aspect ratios evolve under the self-energy expansion. The
theoretical curves have been obtained solving the Castin-Dum
equations without free parameters. (a) Evolution of the $x/z$
aspect ratio. (b) Evolution of the $y/z$ aspect ratio. The data
points are the average of four shots and the error bars represent
their shot to shot variation.} \label{Castdum}
\end{figure}

We use the procedure outlined in section 2 to obtain a BEC in a
normal circular TOP trap and then adiabatically change  the
amplitude of $B_x$, $B_y$ or both to the desired ellipticity E. We
found that the BEC remained condensed during this adiabatic
expansion, during which the RF knife was switched off.
Fig.\ref{BECpic} shows time of flight images of the condensate
after release from the rotating elliptical trap. By releasing the
BEC when the probe beam was aligned with its long or its short
axis the evolution of the x to z and y to z aspect ratios of the
BEC was monitored. The inversion of the shape as the time of
flight increases clearly demonstrates that the expansion of the
cloud is dominated by its self energy. For the data shown in
Fig.\ref{Castdum} $B_x$ remained at its final vaue of $2G$ and
$B_y$ was ramped up to a value of $8G$ in $2s$ giving an
ellipticity of the time-averaged potential of $E=4$ and $e\approx
1.4$. These fields imply a static trap frequency of $70 Hz$. With
a rotation frequency of $30 Hz$, one obtains an effective trap
frequency $(\omega_x^2 -\Omega^2)^{1/2}=63 Hz$. We numerically
solved the Castin-Dum \cite{Castin} equations for the effective
trap frequencies in a non-rotating frame. Fig.\ref{Castdum} shows
the comparison of experimental and theoretical data, which are in
good agreement. This proves that our BEC has survived the rotation
of the elliptical trap and that the Coriolis effects have had
negligible influence.

\section{Conclusions}
We have constructed an anisotropic and rotating TOP trap and shown
that a BEC may be created and transferred to the rotating trap.
The measured properties at low rotation frequencies agree well
with the existing theory. Faster rotation will be a useful tool
for further studies of superfluid properties of a BEC, such as
nucleation and stabilization of vortices. \\

\noindent We gratefully acknowledge fruitful discussions with J.
Anglin, C. Clark, M. Edwards and D. Feder.\\

\noindent This work was supported by the EPSRC and the TMR program
(No. ERB FMRX-CT96-0002). O. Marag\`{o} acknowledges the support
of a Marie Curie Fellowship, TMR program (No. ERB FMBI-CT98-3077).


\newpage

\section*{References}
\begin{thebibliography}{25}

\bibitem{Anderson} M.H. Anderson {\it et al.}, Science {\bf 269}, 198 (1995).
\bibitem{Davis} K.B. Davis {\it et al.}, Phys. Rev. Lett. {\bf 75}, 3969 (1995).
\bibitem{Varenna} Proceedings of the International School of Physics 'Enrico Fermi' edited by
M. Inguscio, S. Stringari and C.E. Wieman, in print (1998).
\bibitem{Jin} D.S. Jin {\it et al.}, Phys. Rev. Lett. {\bf 77}, 420 (1996).
\bibitem{Mewes} M.-O. Mewes {\it et al.}, Phys. Rev. Lett. {\bf 77}, 988 (1996).
\bibitem{Laser} M.-O. Mewes {\it et al.}, Phys. Rev. Lett. {\bf 78}, 582 (1997).
\bibitem{Sodium} E.W. Hagley {\it et al.}, Science {\bf 283}, 1706 (1999).
\bibitem{Component} E.A. Cornett {\it et al.}, J. Low Temp. Phys. {\bf 113}, 151 (1998).
\bibitem{Compo} J. Stenger {\it et al.}, Nature {\bf 396}, 345 (1998).
\bibitem{Interference} M.R. Andrews {\it et al.}, Science {\bf 275}, 637 (1997).
\bibitem{Interfere} B.P. Anderson {\it et al.}, Science {\bf 282}, 1686 (1998).
\bibitem{Theory} D.L. Feder {\it et al.}, Phys. Rev. Lett. {\bf 82}, 4956 (1999).
\bibitem{Stringari} S. Stringari {\it et al.}, Phys. Rev. Lett. {\bf 82}, 4371 (1999).
\bibitem{Fetter} A.A. Svidzinsky {\it et al.}, Phys. Rev. A {\bf 58}, 3168 (1998).
\bibitem{Lundh} E. Lundh {\it et al.}, Phys. Rev. A {\bf 58}, 4816 (1998).
\bibitem{Bolda} E.L. Bolda {\it et al.}, Phys. Rev. Lett. {\bf 81}, 5477 (1998).
\bibitem{Tilley} D.R. Tilley and J.Tilley, "Superfluids and Superconductivity", Adam Hilger, (3rd ed
1991).
\bibitem{Ioffe} D.E. Pritchard {\it et al.}, Phys. Rev. Lett. {\bf 51}, 1336 (1983).
\bibitem{Stir} B.M. Caradoc-Davies {\it et al.}, Phys. Rev. Lett. {\bf 83}, 895 (1999).
\bibitem{Imprinting} R. Dum {\it et al.}, Phys. Rev. Lett. {\bf 80}, 2972 (1998).
\bibitem{Vortex} M.R. Matthews {\it et al.}, Phys. Rev. Lett. {\bf 83}, 2498 (1999).
\bibitem{Angular} D.A. Butts {\it et al.}, Nature {\bf 397}, 327 (1999).
\bibitem{Petrich} W. Petrich {\it et al.}, Phys. Rev. Lett. {\bf 74}, 3352 (1995).
\bibitem{Wilson} J. Martin {\it et al.}, J. Phys. B {\bf32}, 3065 (1999).
\bibitem{Wilson2} A. Wilson {\it et al.}, Appl. Opt. {\bf 37}, 4871 (1998).
\bibitem{double} C.J. Myatt {\it et al.}, Opt. Lett. {\bf 21}, 290 (1996).
\bibitem{Arlt2} J. Arlt {\it et al.}, Opt. Comm. {\bf157}, 303 (1998).
\bibitem{Getter} C. Wieman {\it et al.}, Am. J. Phys. {\bf63}, 317 (1995).
\bibitem{Kasevich} B.P. Anderson {\it et al.}, Phys. Rev. A {\bf 59}, R938 (1999).
\bibitem{Webster} S. Webster, D.Phil. thesis, Oxford University (to be published).
\bibitem{Arlt}  J. Arlt {\it et al.}, J. Phys. B {\bf 31}, L321 (1998).
\bibitem{Dalibard} M. Arndt {\it et al.}, Phys. Rev. Lett. {\bf 79}, 625 (1997).
\bibitem{Lamb} H. Lamb, "Dynamics", Cambridge University Press (2nd ed 1923).
\bibitem{Castin} Y. Castin {\it et al.}, Phys. Rev. Lett. {\bf 77}, 5315 (1996).

\end{thebibliography}
\end{document}